\documentclass[aps,prl,twocolumn,showpacs]{revtex4}
\usepackage{graphicx}
\usepackage{bm}
\usepackage{amsmath}
\usepackage{amssymb}

\providecommand{\abs}[1]{\left\lvert#1\right\rvert}

\providecommand{\bra}[1]{\langle #1 \rvert}
\providecommand{\ket}[1]{\lvert #1 \rangle}

\begin{document}

\title{Playing a quantum game on polarization vortices}
 
\author{A.~R.~C.~Pinheiro$^{1}$, C.~E.~R. Souza$^{1}$, D.~P.~Caetano$^{2}$, 
J.~A.~O.~Huguenin$^{3}$, A.~G.~M.~Schmidt$^{3}$, and A.~Z.~Khoury$^{1}$}

\affiliation{$^{1}$
Instituto de F\'\i sica, Universidade Federal Fluminense,
24210-346 Niter\'oi - RJ, Brasil.}
\affiliation{$^{2}$Escola de Engenharia Industrial Metal\'urgica, Universidade Federal Fluminense, 
27255-125 Volta Redonda - RJ, Brazil}
\affiliation{$^{3}$Instituto de Ci\^encias Exatas, Universidade Federal Fluminense, 
27213-415 Volta Redonda - RJ, Brazil}

\begin{abstract}
The quantum mechanical approach to the well known prisoners dilemma, one of the 
basic examples to illustrate the concepts of Game Theory, is implemented with 
a classical optical resource, nonquantum entanglement between spin and orbital 
degrees of freedom of laser modes. The concept of 
entanglement is crucial in the quantum version of the game, which brings novel 
features with a richer universe of strategies. As we show, this richness can be 
achieved in a quite unexpected context, namely that of paraxial spin-orbit modes 
in classical optics. 

\end{abstract}
\vskip2pc 
 
\maketitle

Numerous quantum information protocols rely on entanglement, a property usually 
attributed to composite quantum systems that cannot be described by a tensor 
product state vector. Quantum cryptography and teleportation are popular 
examples of protocols relying on quantum entanglement. While this property is 
considerably sensitive to local measurements performed in each party of the 
composite system, it is unaffected by local unitary operations. Teleportation 
protocols, for example, are achieved by classical communication followed by 
local unitary operations and measurements. This framework is where Quantum 
Mechanics meets an important area of applied Mathematics, the Game theory, 
a powerful tool for decision making \cite{meyer,survey}. 
Here, two or more agents (players) take 
their decisions by acting on a quantum system with unitary operations. 
These decisions or conflict situations can be as simple as tossing a 
coin\cite{meyer} or rather involved like the so-called minority game 
\cite{challet}, where one models a competition among several players 
for a limited resource. In this sense games can be 
cooperative or non-cooperative like the prisoners dilemma, and with 
complete (incomplete) information where one player knows (does not 
know) all strategies his opponent can choose. 
Quantum versions of this game was realized 
experimentally both with Nuclear Magnetic Resonance \cite{nmr} and 
entangled photon pairs \cite{schmid}. 

Although frequently attributed to quantum systems, entanglement has been recently 
identified in classical optics as the coherent superposition of paraxial modes 
with orthogonal spatial profiles and orthogonal polarizations. We shall 
refer to such superpositions as \textit{spin-orbit modes}. These modes 
were used in our group as a classical optical resource to investigate 
the topological phase acquired by an entangled state following a cyclic 
evolution under local unitary operations \cite{fasetopologica} and to 
demonstrate alignment free BB84 quantum cryptography ref.\cite{crypto}. 
Also, a spin-orbit Bell 
inequality has been investigated both in the quantum \cite{bellchines} and 
classical \cite{bellnosso} domains. An important tool for 
spin-orbit coupling was used in ref.\cite{quplate}
to exchange quantum information between these two degrees of freedom. 
In a recent work \cite{simon} the term 
\textit{nonquantum entanglement} has been coined to designate the spin-orbit 
inseparability of paraxial modes in connection with the Mueller matrices 
employed in polarization optics. In this work we demonstrate how this 
kind of nonquantum entanglement can be used to evaluate the performance 
of quantum strategies in a classical example of game theory, the 
prisoners dilemma.  

A laser beam propagating along the $z$ direction is usually described by its 
polarization unit vector $\hat{e}$ and its spatial mode $\psi(\mathbf{r})$. 
The spatial modes in rectangular coordinates are Hermite-Gaussian (HG) solutions 
of the paraxial wave equation described in many text books \cite{yariv}. 
The subspace of first order spatial modes has a qubit structure similar to the 
polarization mode space, where HG modes along different orientations play the role 
of linear polarizations and the Laguerre-Gaussian (LG) modes are equivalent to 
circular polarization \cite{miles}. By combining the two mode spaces, a general 
spin-orbit mode can be written as
\begin{eqnarray}
\mathbf{\Psi}(\mathbf{r})&=&
\alpha\,\psi_{h}(\mathbf{r})\,\hat{e}_{H} + \beta\,\psi_{v}(\mathbf{r})\,\hat{e}_{H} 
\nonumber \\
&+&\gamma\,\psi_{h}(\mathbf{r})\,\hat{e}_{V} + \delta\,\psi_{v}(\mathbf{r})\,\hat{e}_{V}\;,
\label{spmode}
\end{eqnarray}
where $\hat{e}_{H(V)}$ are linear polarization unit vectors along the horizontal 
(vertical) directions, and $\psi_{h(v)}(\mathbf{r})$ are HG spatial modes 
along these same directions. In analogy to the usual entanglement measure used for 
bipartite quantum states \cite{wooters}, we can define the spin-orbit mode 
concurrence, which for eq.(\ref{spmode}) is $C=\abs{\alpha\delta-\beta\gamma}$.
We shall refer to the spin-orbit modes with maximal concurrence $C=1$ as 
maximally entangled modes. In these modes neither the polarization 
nor the spatial profile is well defined, in fact they correspond to polarization 
vortices which have been extensively studied due to their potential applications 
to high resolution microscopy \cite{polvortex1, polvortex2}. 
In this work we demonstrate another 
appealing feature in the unexpected context of a quantum game.  
An example of such entangled spin-orbit mode is
\begin{equation}
\mathbf{\Psi}_0(\mathbf{r}) =
\frac{\psi_{h}(\mathbf{r})\,\hat{e}_{H} + i\,\psi_{v}(\mathbf{r})\,\hat{e}_{V}}
{\sqrt{2}}\;.
\label{psi0}
\end{equation}

We use this mode to implement a classical optical version of the well known 
prisoners dilemma, in which two players, Alice and Bob, accused of a 
felony, have to decide whether they cooperate ($C$) or defeat ($D$) each other. In 
classical game theory, each agent decision is represented by one bit 
of information with possible states $C$ and $D$. Depending on their decision, 
a reduced penalty may be applied to each one of them. The penalty 
reduction is the payoff each player gets from their combined decisions. 
The penalty reductions for both players are shown in table \ref{redtab} 
for all possible strategies adopted by Alice (rows) and Bob (columns). 
\begin{table}[h]
	\centering
		\begin{tabular}{|c|c|c|}
\hline
$(R_A,R_B)$ & $C$ & $D$\\ \hline
$C$  & (3,3) & (0,5)\\ \hline
$D$  & (5,0) & (1,1)\\ \hline			
		\end{tabular}
	\caption{Penalty reduction table}
	\label{redtab}
\end{table}
From the table we see that both players are tempted to choose $D$, although 
their added reduction would be maximized by $CC$. Here comes the dilemma, 
the players are isolated without the permission to negotiate. 
Each player is left to his own and has to decide whether to defeat or cooperate 
with the other, a bad choice may cost his freedom. For a cooperative game, 
players would choose strategies which maximize both payoffs, i.e., they would 
search for Pareto optimal strategies; on the other hand, since prisoners dilemma 
is a non-cooperative game, each player will try to maximize solely his own payoff, 
i.e., the intelligent choice is the Nash equilibrium $DD$. At this point, the 
concepts of Game Theory are due. Suppose this situation is repeated many times 
and the players adopt probabilistic strategies, that is, they randomly choose 
between $C$ and $D$ with prestablished probabilities. Then, the payoff function 
of each player is given by the average penalty reduction obtained:
\begin{equation}
\$_j=\sum_{m,n=C,D}\,p\,(m,n)\,R_{j}(m,n)\;,
\label{payoff}
\end{equation}
where $j=A,B$ and $p\,(m,n)=p_A(m)p_B(n)$ is the joint probability that Alice chooses $m$ and 
Bob chooses $n$. In the classical approach to the problem, one shows that the payoff 
as a function of $p_A(m)$ and $p_B(n)$ has an absolute minimum at $p_A(D)=p_B(D)=1$, that is 
the best the players can do is to defeat, as a consequence of the severe cost brought 
by a possible betrayment.  

In ref.\cite{eisert} Eisert proposes an ingenuous alternative to the classical approach, 
employing the concept of quantum entanglement. Briefly, in this approach the prisoners share 
a pair of entangled qubits and rather than making a definite $C$ or $D$ statement, they 
are allowed to perform single qubit unitary operations (strategies), each one on the qubit 
in his possession. The entangled two-qubit state is prepared by a nonlocal operation 
\begin{equation}
U = \left[
\begin{matrix}
1&0&0&i\\
0&1&i&0\\
0&i&1&0\\
i&0&0&1
\end{matrix}
\right]
\label{U}
\end{equation}
acting on an initial state $\ket{CC}$, so that 
$U\ket{CC}=\left(\ket{CC}+i\ket{DD}\right)/\sqrt{2}\,$. 
After each player has applied his own strategy $U_j$ ($j=A,B$), 
the qubits are nonlocally 
operated with $U^{\dagger}$ and separately 
measured. The payoff function (\ref{payoff}) is evaluated with the probabilities 
$p(m,n)=\bra{m\,n}U^{\dagger}\left(U_A\otimes U_B\right)U\ket{CC}$ ($m,n=C,D$)
associated with the two possible outcomes ($C$ or $D$) in each qubit. Therefore, the 
strategy space is much larger in this quantum approach, it corresponds to the space 
of $SU(2)\otimes SU(2)$ matrices, apart from irrelevant global phases. This quantum 
version has been implemented experimentally with quantum correlated photon pairs 
generated by parametric down conversion \cite{zeilinger}. 

%
\begin{figure}[h]
\begin{center} 
\includegraphics[scale=0.35]{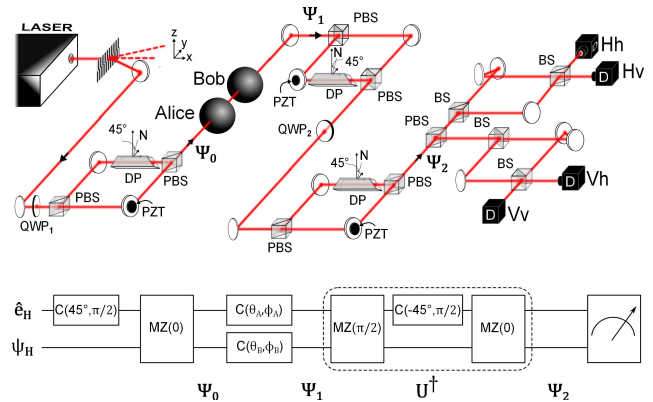}
\end{center} 
\caption{Experimental setup.}
\label{setup}
\end{figure}
%

We now demonstrate the implementation of the quantum game strategies with the 
nonquantum entanglement provided by the spin-orbit classical mode (\ref{psi0}). 
In the game language, we shall make the identification $H\equiv C$ and $V\equiv D$. 
The experimental setup is shown in fig.(\ref{setup}). 
The TEM$_{00}$ output of a He-Ne laser is diffracted by a hologram 
that generates an HG mode on the first diffraction order, so that the 
initial spin-orbit mode is $\psi_h(\mathbf{r})\,\hat{e}_H$. This mode is first sent 
to a quarter wave plate (QWP) rotated at $45^o$, which makes the transformation 
$\hat{e}_H\to\left(\hat{e}_H + i\,\hat{e}_V\right)/\sqrt{2}$, and then to 
a Mach-Zehnder (MZ) interferometer with input and output polarizing beam splitters (PBS). 
A Dove prism (DP) is inserted in one arm of the MZ 
interferometer, and the relative phase $\phi$ between the two arms is controlled by a 
piezoelectric transducer (PZT). The DP oriented at $45^o$ makes the transformation 
$\psi_h\rightarrow\psi_v$, so that this MZ interferometer coherently superposes modes 
$\psi_v(\mathbf{r})\,\hat{e}_V$ and $\psi_h(\mathbf{r})\,\hat{e}_H$ at its output 
with an adjustable phase $\phi$. In the basis 
$\{\psi_h(\mathbf{r})\,\hat{e}_H, \psi_v(\mathbf{r})\,\hat{e}_H, 
\psi_h(\mathbf{r})\,\hat{e}_V, \psi_v(\mathbf{r})\,\hat{e}_V\}$, 
the matrix representation of the MZ transformation is:
\begin{equation}
MZ(\phi) = \left[
\begin{matrix}
1&0&0&0\\
0&-1&0&0\\
0&0&0&-e^{i\phi}\\
0&0&e^{i\phi}&0
\end{matrix}
\right]\;,
\label{MZ}
\end{equation}
so that after passing through the QWP and the balanced interferometer ($\phi=0$), 
the beam is prepared in mode (\ref{psi0}). This mode is the object on which the 
players will implement their strategies. 

In order to continue our experimental description, it is essential to define 
the mode converter operators which will be used to implement the players 
strategies and $U^{\dagger}$. We will be dealing with either polarization 
(wave plates) or spatial (DP and cylindrical lenses) mode converters, that is, 
elements acting on one degree of freedom only. We shall inform rotation 
angles in degrees and phase retardations in radians for immediate identification 
of the physical meaning all over the experimental description. 
When oriented along horizontal and vertical directions, they introduce a 
retardation phase $\phi$ between $H$ and $V$ modes. A mode converter rotated 
by the angle $\theta$ is described by the $SU(2)$ matrix:
\begin{equation}
C(\theta,\phi) = \left[
\begin{matrix}
\cos\frac{\phi}{2}+i\sin\frac{\phi}{2}\cos 2\theta & i\sin\frac{\phi}{2}\sin 2\theta\\
i\sin\frac{\phi}{2}\sin 2\theta & \cos\frac{\phi}{2}-i\sin\frac{\phi}{2}\cos 2\theta
\end{matrix}
\right]\;.
\label{C}
\end{equation}
For example, quarter wave plates correspond to $\phi=\pi/2$ and 
half wave plates (HWP) to $\phi=\pi$. Spatial mode converters can be made 
with cylindrical lenses \cite{converter} for variable retardation $\phi$, 
or the DP for $\phi=\pi$. 
Now, Alice is equipped with polarization elements and realizes strategies of 
the kind $C(\theta_A,\phi_A)$, whereas Bob is equipped with DPs and cylindrical 
lenses, and his strategies are $C(\theta_B,\phi_B)$. After the players have 
made their choices, the spin-orbit mode of the laser beam is:
\begin{equation}
\mathbf{\Psi}_1(\mathbf{r}) = \left[C(\theta_A,\phi_A)\otimes C(\theta_B,\phi_B)\right]
\mathbf{\Psi}_0(\mathbf{r})
\;.
\label{psi1}
\end{equation}

The mode converters are also used together with $MZ(\phi)$ to implement $U^{\dagger}$. 
Indeed, one can easily show that:
\begin{equation}
U^{\dagger}=MZ(0)\,\left[C(-45^o,\pi/2)\otimes I\right]\,MZ(\pi/2)\;,
\label{Udagger}
\end{equation}
which physically means that, after passing through the players 
strategies, the beam is sent through a MZ interferometer with a $\pi/2$ 
phase shift, a quarter wave plate (QWP) rotated at $-45^o$, and another 
MZ interferometer with balanced arms ($\phi=0$), where it gets  
transformed to mode 
\begin{equation}
\mathbf{\Psi}_2(\mathbf{r})=U^{\dagger}\mathbf{\Psi}_1(\mathbf{r})
=\sum_{m,n} c_{mn}\,\psi_m(\mathbf{r})\,\hat{e}_n\;,
\label{psi2}
\end{equation}
where $m=h,v$ and $n=H,V$.

After this transformation, we arrive at the measurement stage where the 
probabilities $p(m,n)$ used in the payoff function eq.(\ref{payoff}) are obtained. 
In our setup, these probabilities are given by the projected intensities: 
\begin{eqnarray}
p(m,n)=\abs{c_{mn}}^2=\abs{\int d^2\mathbf{r}\,\,\psi_m^*(\mathbf{r})
\left[ \hat{e}_n^* \cdot\mathbf{\Psi}_2(\mathbf{r})\right]}^2
\;,
\label{pmn}
\end{eqnarray}
which, in fact, correspond to photodetection probabilities when the beam 
is attenuated down to the single photon regime. In order to measure the 
projected intensities, the $\mathbf{\Psi}_2(\mathbf{r})$ mode is first 
sent to a PBS where polarization projection is performed. Then, each 
PBS output is sent to a balanced Mach-Zehnder interferometer with an 
additional mirror in one arm (MZIM), where spatial mode projection 
occurs \cite{mzim}. The projected intensities are measured at the four 
outputs of the two MZIMs either with a CCD camera or with photodetectors. 
The payoff function is evaluated 
with the intensities measured with four photodetectors. The background noise 
is subtracted and the partial intensities are then normalized to the total 
intensity so that $p(m,n)\equiv I_{m,n}/I_{TOT}$. 

%
\begin{figure}[h]
\begin{center} 
\includegraphics[scale=0.3]{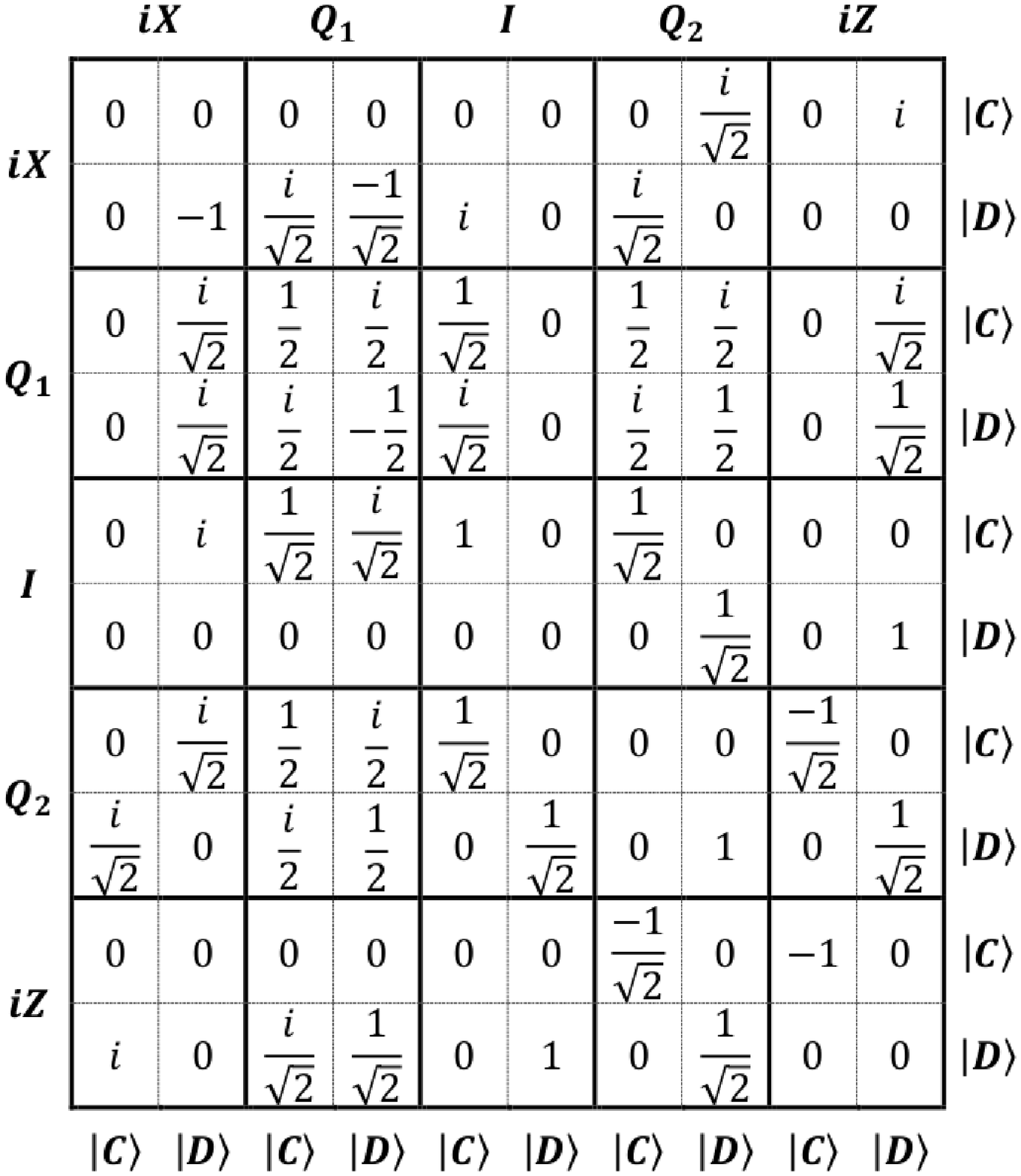}
\includegraphics[scale=0.3]{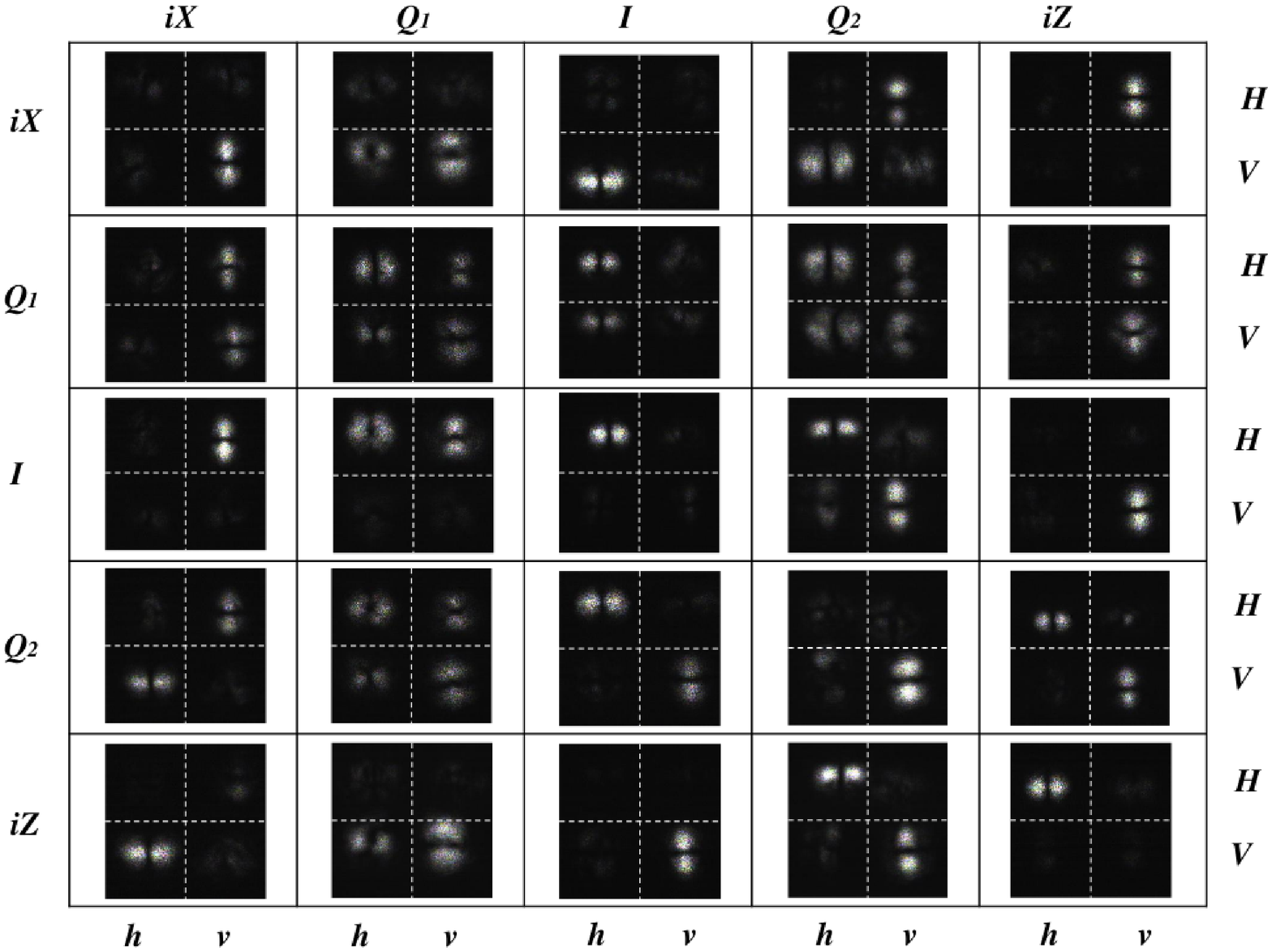}
\end{center} 
\caption{a) Coefficients table of the final mode in the computational basis. 
b) Images of the corresponding output ports.}
\label{tables}
\end{figure}
%

\begin{figure}[h]
\begin{center} 
\includegraphics[scale=0.25]{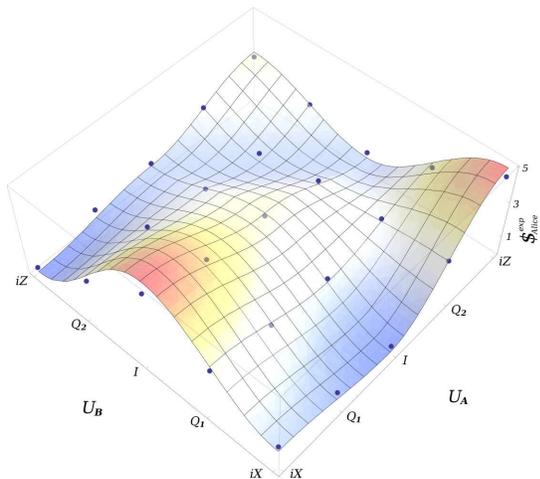}
\end{center} 
\caption{Alice payoff as a function of the strategies parameters $(\theta_A,\phi_A)$ 
and $(\theta_B,\phi_B)$. Dots correspond to the experimental values obtained with 
the intensity measurements.}
\label{figpayoff}
\end{figure}

For each player, we implemented five different strategies: $iX\equiv C(45^o,\pi)$, 
$Q_1\equiv C(45^o,\pi/2)$, $I\equiv C(\theta,0)$, $Q_2\equiv C(0,\pi/2)$, and 
$iZ\equiv C(0,\pi)$. In this notation $X$ and $Z$ are the usual Pauli matrices. 
Strategies $I$ and 
$iX$ are equivalent to the classical ones where the players 
can only cooperate or defeat, while the other strategies are intrinsically 
quantum mechanical since they involve rotations and phase retardations 
not available in the classical scenario.
A table with the coefficients $c_{mn}$ resulting from  
these strategies is shown in fig.(\ref{tables}a). Also, the images obtained with the CCD 
camera are displayed in the same table format in fig.(\ref{tables}b). As expected, only those 
output ports corresponding to nonzero coefficients are illuminated. In this sense, 
the qualitative agreement between the two tables is clear. 
We have also evaluated Alice's payoff as a function of the 
strategy parameters $(\theta_A,\phi_A)$ and $(\theta_B,\phi_B)$ in 
the domain $(\theta=0,0\leq \phi\leq\pi)$ and $(\theta=45^o,0\leq \phi\leq\pi)\,$.
The analytical result is shown in fig.(\ref{figpayoff}) together with the 
points corresponding to all possible combinations of the experimental strategies. 
The experimental values were obtained from the intensity measurements, where the 
relative intensities play the role of measurement probabilities. 
We observe, from both theoretical and experimental results, that the 
quantum move $U_A=U_B=iZ$ proposed originally in \cite{eisert} dominates all 
classical ones and it is also a Nash equilibrium with a better outcome than 
$U_A=U_B=iX$ since no player can improve his respective payoff by changing 
unilaterally his strategy. Note also that if Bob could choose only between 
classical $U_B=I$ or $U_B=iX$, and Alice keep $U_A=iZ$ then it would be better 
for him to cooperate since his payoff would be increased.

%
%

In conclusion, we have used the concept of nonquantum entanglement to 
implement a Game Theory task in the context of the well known prisoners 
dilemma. The advantages offered by the quantum mechanical approach could 
be realized in the classical optics framework. Nonseparable spin-orbit modes 
corresponding to polarization vortices were used. 
This implementation opens promising perspectives regarding 
potential applications of nonquantum entanglement to the investigation of 
quantum information protocols.

\section*{Acknowledgments}
Funding was provided by Coordena\c c\~{a}o de Aperfei\c coamento de 
Pessoal de N\'\i vel Superior (CAPES), Funda\c c\~{a}o de Amparo \`{a} 
Pesquisa do Estado do Rio de Janeiro (FAPERJ-BR), and Instituto Nacional 
de Ci\^encia e Tecnologia de Informa\c c\~ao Qu\^antica (INCT-CNPq).

\end{document}